\documentclass[a4paper,12pt,nonacm]{acmart}
\usepackage{graphicx}
\usepackage{tikz}
\usepackage{geometry}
\usepackage{xcolor}
\usepackage[T1]{fontenc}
\usepackage{tgbonum}
\usepackage{lmodern}
\usepackage{tgtermes}
\usepackage{tgpagella}
\usepackage{tgschola}
\usepackage{mathptmx}
\usepackage{utopia}
\usepackage{palatino}
\usepackage{bookman}
\usepackage{charter}
\usepackage{multirow}
\usetikzlibrary{positioning, shapes, arrows, fit}
\definecolor{offwhite}{RGB}{220,220,220}

\newcommand{\papertitle}{What Are The Risks of Living in a GenAI Synthetic Reality? The Generative AI Paradox}
\newcommand{\paperauthors}{Emilio Ferrara}
\newcommand{\paperaffiliation}{University of Southern California} % DO NOT CHANGE

\begin{document}
%\title{}
%\maketitle

% Cover Page
\begin{titlepage}
    \begin{tikzpicture}[remember picture, overlay]
        % Provide bounding box if LaTeX cannot determine it
        \node[anchor=north west, inner sep=0] at (current page.north west) {
            \includegraphics[width=\paperwidth, height=\paperheight, trim=275 375 275 375]{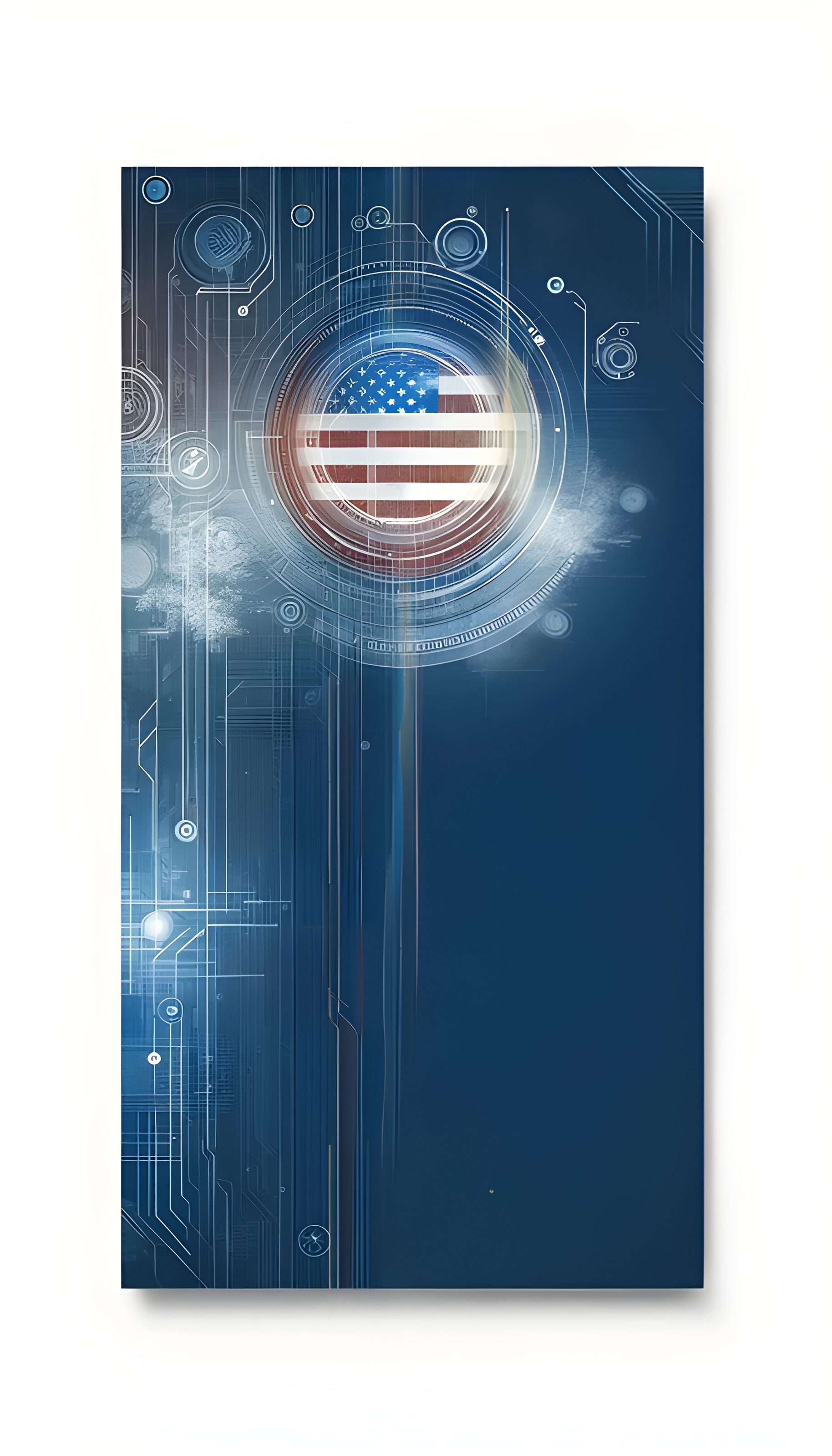}
        };
        \node[anchor=center, yshift=-9cm] at (current page.center) {
            \begin{minipage}{\textwidth}
                \raggedleft
                \color{offwhite}
                % Title
                {\Huge \bfseries \fontfamily{qtm}\selectfont The 2024 Election Integrity Initiative }
                
                \vspace{1.5cm}
                
                % Paper title
                {\LARGE \fontfamily{qtm}\selectfont \papertitle}
                
                \vspace{1.5cm}
                
                % Authors
                {\Large \fontfamily{qtm}\selectfont \paperauthors}
                
                \vspace{1cm}
                
                % Institution
                {\Large \fontfamily{qtm}\selectfont \paperaffiliation}
                
                \vfill
                
                % Working Paper Number
                {\Large \fontfamily{qtm}\selectfont HUMANS Lab -- Working Paper No. 2024.2}
            \end{minipage}
        };
    \end{tikzpicture}
\end{titlepage}

% Following content (example)
\noindent{\LARGE \fontfamily{qtm}\selectfont \papertitle}

\vspace{0.5cm}

\noindent{\large \fontfamily{qtm}\selectfont \paperauthors}

\noindent{\large \fontfamily{qtm}\selectfont \textit{\paperaffiliation}}

% \section*{Abstract}
% Your abstract content here

\begin{figure}[b]
  \centering
  \begin{tabular}[t]{cc}
    \multirow{2}{*}[6.9cm]{\includegraphics[height=.81\linewidth]{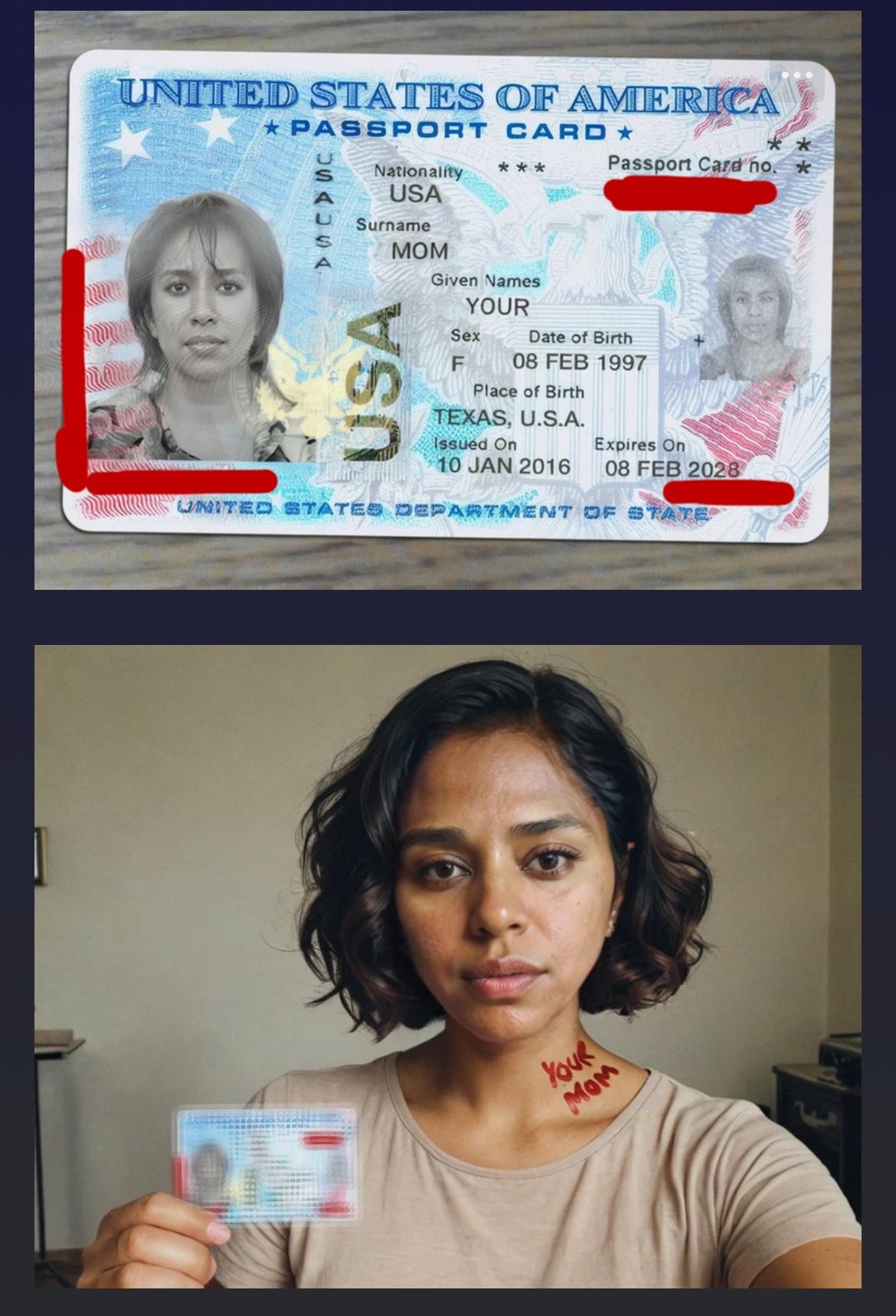}} & 
    \includegraphics[height=.4\linewidth]{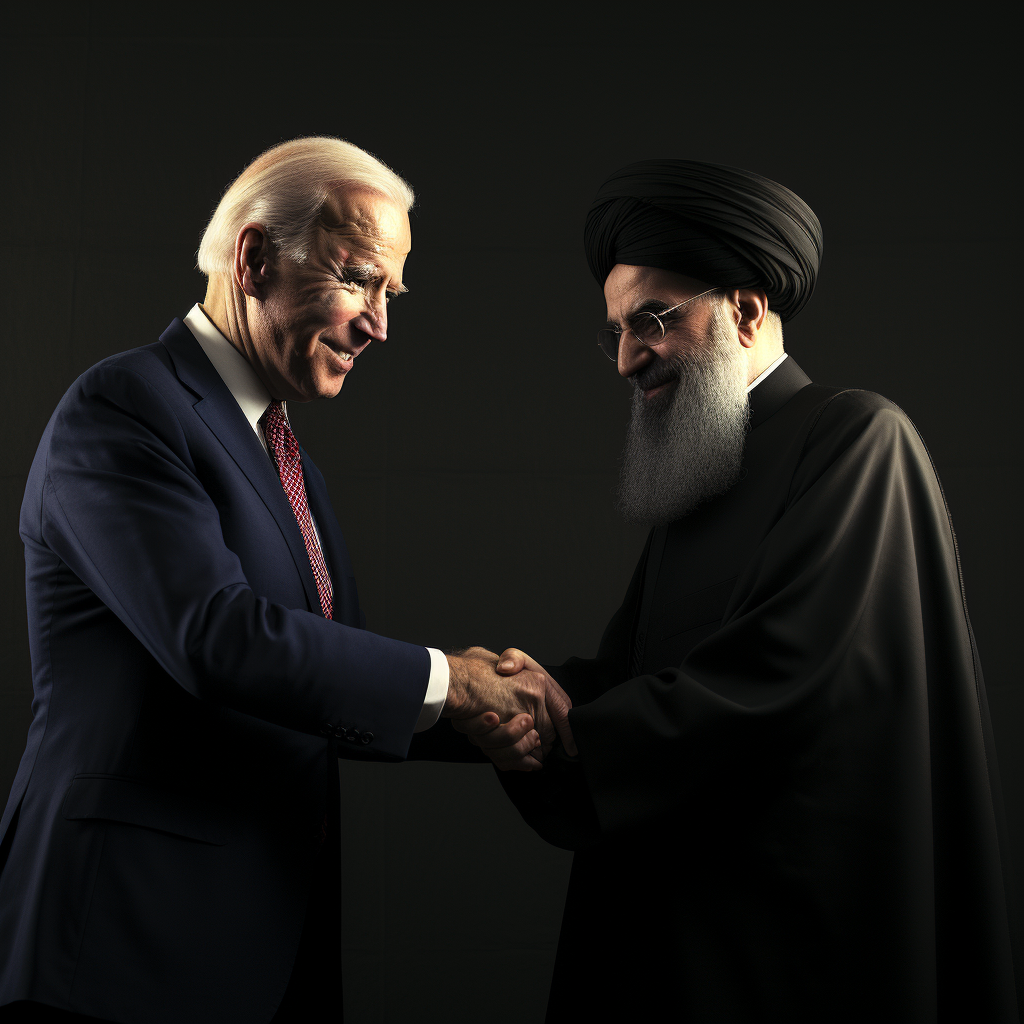} \\
    & \includegraphics[height=.4\linewidth]{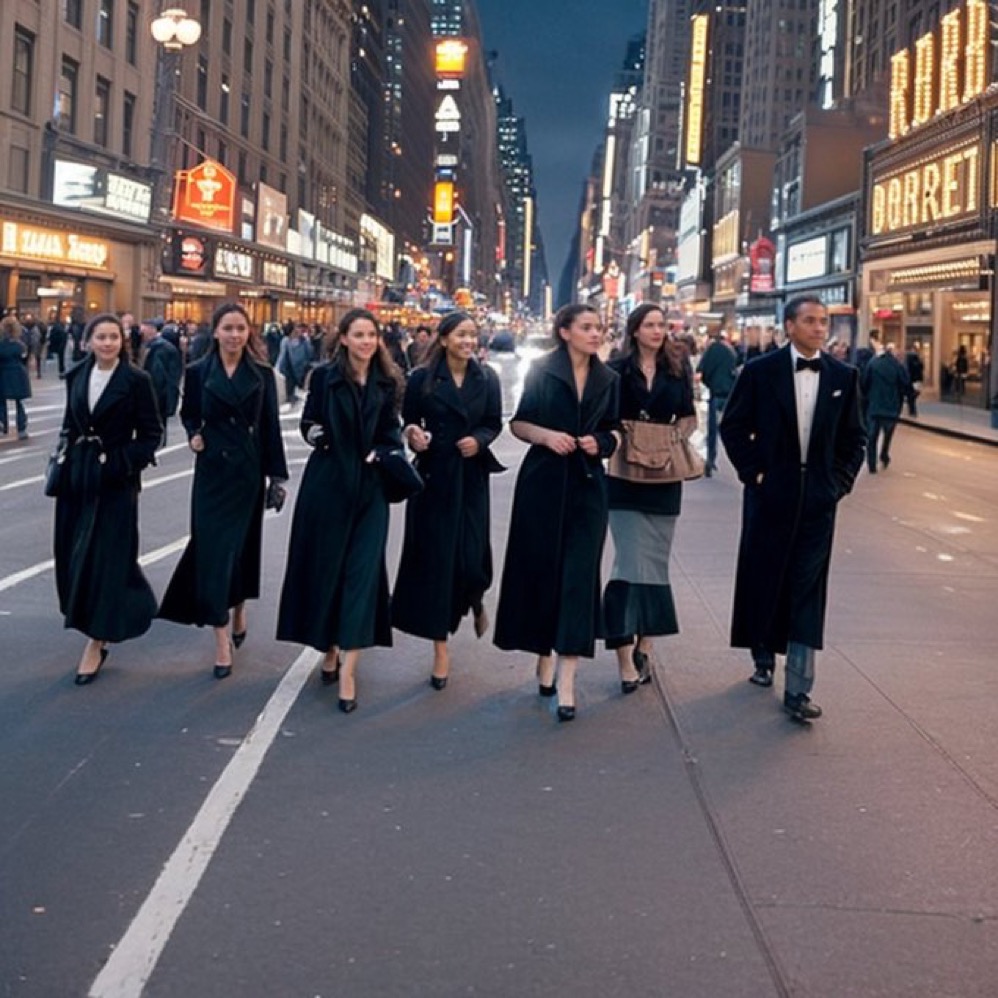} \\
  \end{tabular}
  \caption{(Left) In January 2024, the \textsl{r/StableDiffusion} community on Reddit demonstrated a proof-of-concept workflow to synthetically generate proofs of identity. (Top Right) GenAI can produce lifelike depictions of never-occurred events (MJv5 prompt: "\textit{president biden and supreme leader of iran shaking hands}"). (Bottom Right) Subliminal messages in generated content (optical illusion reads \textit{OBEY}).}
  \label{fig:teaser}
\end{figure}

\section*{Introduction}
Generative AI (GenAI) technologies possess unprecedented potential to reshape our world and our perception of reality. These technologies can amplify traditionally human-centered capabilities, such as creativity and complex problem-solving in socio-technical contexts.\footnote{State of California, \textit{Benefits and Risks of Generative Artificial Intelligence}, Report, State of California, November 2023.} By fostering human-AI collaboration, GenAI could enhance productivity, dismantle communication barriers across abilities and cultures, and drive innovation on a global scale.\footnote{Tojin T. Eapen, Daniel J. Finkenstadt, Josh Folk, and Lokesh Venkataswamy, \href{https://hbr.org/2023/07/how-generative-ai-can-augment-human-creativity}{How Generative AI Can Augment Human Creativity}, Harvard Business Review, July 2023.} 

Yet, experts and the public are deeply divided on the implications of GenAI. Concerns range from issues like copyright infringement and the rights of creators whose work trains these models without explicit consent,\footnote{Gil Appel, Juliana Neelbauer, and David A. Schweidel, \href{https://hbr.org/2023/04/generative-ai-has-an-intellectual-property-problem}{Generative AI Has an Intellectual Property Problem}, Harvard Business Review, April 2023.} to the conditions of those employed to annotate vast datasets.\footnote{Billy Perrigo, \href{https://time.com/6247678/openai-chatgpt-kenya-workers/}{OpenAI's ChatGPT Has Kenya Workers}, Time, January 2023.} Accordingly, new laws and regulatory frameworks are emerging to address these unique challenges.\footnote{Christopher T. Zirpoli, \textit{Generative Artificial Intelligence and Copyright Law}, Congressional Research Service, September 2023.} Others point to broader issues, such as economic disruptions from automation and the potential impact on labor markets. Although history suggests that society can adapt to such technological upheavals, the scale and complexity of GenAI's impact warrant careful scrutiny.

This paper, however, highlights a subtler, yet potentially more perilous risk of GenAI: the creation of \textit{personalized synthetic realities}. GenAI could enable individuals to experience a reality customized to personal desires or shaped by external influences, effectively creating a "filtered" worldview unique to each person. Such personalized synthetic realities could distort how people perceive and interact with the world, leading to a fragmented understanding of shared truths. This \textit{Viewpoint} seeks to raise awareness about these profound and multifaceted risks, emphasizing the potential of GenAI to fundamentally alter the very fabric of our collective reality.

\subsection*{A Taxonomy of GenAI Risks and Harms}
At the heart of these concerns is a taxonomy of GenAI risks and harms, as proposed in \cite{ferrara2024genai}. This taxonomy not only categorizes the risks associated with GenAI but also underscores the critical need for proactive strategies by linking specific intents (\textit{dishonesty, propaganda, deception}) to the types of harm they are likely to produce.

\paragraph{Personal Loss}
This category encompasses \textit{harm to individuals}, including threats such as identity theft and privacy invasion. GenAI’s capability to synthesize highly realistic but false representations of people creates significant personal risks, including breaches of private information, defamation, and a growing erosion of public trust \cite{menczer2023addressing, seymour2023beyond}.

\paragraph{Financial and Economic Damage}
Beyond individual harm, GenAI poses threats to societal and economic stability. This category includes risks like GenAI-driven financial fraud and the potential destabilization of markets through the spread of misinformation, highlighting significant economic vulnerabilities \cite{mazurczyk2023disinformation}.

\paragraph{Information Manipulation}
This dimension addresses GenAI’s capacity to construct false yet persuasive narratives, which threatens the foundations of democratic societies in our increasingly information-saturated environment. The ability to manipulate information at scale raises concerns about the future integrity of public discourse.

\paragraph{Socio-technical and Infrastructural Risks}
Finally, GenAI introduces risks at the socio-technical and infrastructural level, with the potential for catastrophic systemic failures. For example, platforms could intentionally manipulate user emotions or worldviews, while governments might exploit GenAI for hyper-targeted surveillance and censorship, effectively transforming information into a tool of totalitarian control.

\section*{What You Can't Tell Apart Can Harm You}
One might argue, with some merit, that the harms outlined in our taxonomy are not uniquely enabled by GenAI; after all, misinformation and deception have existed long before the digital age. For decades, mass spam has plagued email, while false news, digitally altered images, and even fabricated videos have had considerable influence on public discourse around politics, health, and more. Immersive video games and alternative realities have also evolved, offering increasingly engaging, albeit fictional, experiences.

However, GenAI introduces a set of unique risks that intensify these issues in unprecedented ways. Here are some key challenges specific to GenAI technologies:

\begin{itemize}
    \item \textit{Cost and Commoditization}: GenAI significantly lowers the barriers to creating realistic content, democratizing the process and enabling individuals or groups without specialized skills to generate convincing synthetic media. This accessibility broadens the reach of these technologies, which can be used for both benign and malicious purposes.

    \item \textit{Scale and Mass Production}: GenAI’s scalability facilitates the mass production of customized content, allowing for the rapid and targeted dissemination of misinformation. This capability enables the manipulation of public opinion, election interference, and destabilization of democratic processes on an unprecedented scale \cite{ferrara2024charting}.

    \item \textit{Customization for Malicious Use}: The open-source nature of many GenAI models enables the creation of custom-tailored tools for nefarious purposes. Even if commercially available GenAI tools are regulated, the low-cost development of malicious custom models remains feasible, raising significant concerns.

    \item \textit{Hyper-targeted Attacks}: GenAI enables the creation of highly personalized misinformation campaigns, scams, and other forms of digital manipulation, specifically targeting individuals or groups. Such hyper-targeted attacks risk undermining trust and cohesion within communities, and by extension, society as a whole.

    \item \textit{Challenges in Detection and Watermarking}: Detecting GenAI-generated content remains a significant technological hurdle. While digital watermarking and forensic methods are in development, the rapid evolution of GenAI outpaces these efforts, creating a continual arms race between creation and detection tools and complicating efforts to preserve content authenticity.

    \item \textit{Eroding Trust in Information Sources}: As GenAI content becomes harder to distinguish from human-created content, public trust in media, institutions, and interpersonal communication is increasingly at risk. This erosion of trust may lead to widespread skepticism and cynicism, making it more challenging to address societal issues based on factual evidence.

    \item \textit{Realism and Blurring of Boundaries}: The hyper-realism achievable with GenAI-generated content blurs the line between real and synthetic worlds. This ambiguity presents challenges across fields such as journalism and legal evidence, potentially leading to a society-wide distrust of digital media. Over time, GenAI could foster alternative synthetic realities, customized to individual preferences, potentially resulting in mass escapism and social isolation.
\end{itemize}

In contrast to earlier technologies like photoshopping, the ease, speed, and sophistication of GenAI in generating synthetic realities is unparalleled. This shift necessitates a fundamental reevaluation of how we interact with and assess the authenticity of digital content.

\subsection*{Implications of GenAI Synthetic Realities}
The risks posed by GenAI misuse reach beyond technological concerns, permeating social, ethical, and moral dimensions \cite{kobis2021bad, schramowski2022large, ferrara2016rise, ferrara2019history}. For instance, in January 2024, a Reddit community showcased a GenAI workflow that generated false proofs of identity (\textit{cf.,} Figure 1, Left). Given that most modern security protocols rely on identity verification steps, GenAI’s capacity to produce hyper-realistic personas and documents has significant implications for the integrity of these systems.

The potential to fabricate or "document" events that never occurred (see Figure 2, Top Right) or to embed synthetic evidence within legitimate content could be devastating for law enforcement, democratic institutions, and society as a whole. In cybersecurity, GenAI-enhanced cyberattacks introduce new challenges to digital infrastructure and data protection, with applications in cyber warfare and espionage by nation-states.

Another concerning dimension is the deliberate manipulation of perceived reality via GenAI: corrupted GenAI tools could subtly influence social behaviors and dynamics through subliminal messaging (\textit{cf.,} Figure 3, Bottom Right). This can exacerbate biases, reinforce stereotypes, deepen echo chambers, and contribute to increased polarization and discrimination \cite{caliskan2017semantics, baeza2018bias}. Additionally, governments or corporations with misaligned incentives might exploit GenAI for totalitarian information control or to deepen societal alienation.

Addressing these challenges requires a coordinated response from policymakers, technologists, ethicists, and the public. There is an urgent need for ethical frameworks, transparent practices, and responsible governance to balance the benefits and risks of GenAI. Public awareness and education are essential, alongside policies and regulations focused on privacy, security, and the ethical use of GenAI, to protect societal interests and preserve social integrity.

In the end, these risks reveal GenAI’s profound paradox: society may collectively adopt the assumption that digital content is inherently synthetic, or “fake,” while only lived or directly witnessed experiences are regarded as “real.”

\section*{Acknowledgements}
This work was supported in part by DARPA (contract \#HR001121C0169). The funders had no role
in study design, data collection and
analysis, decision to publish, or preparation of the manuscript.

\section*{About the Team}
The 2024 Election Integrity Initiative is led by Emilio Ferrara and Luca Luceri and carried out by a collective of USC students and volunteers whose contributions are instrumental to enable these studies. The author is grateful to the following HUMANS Lab's members for their tireless efforts on this project: Ashwin Balasubramanian, Leonardo Blas, Charles 'Duke' Bickham, Keith Burghardt, Sneha Chawan, Vishal Reddy Chintham, Eun Cheol Choi, Srilatha Dama, Priyanka Dey, Isabel Epistelomogi, Saborni Kundu, Grace Li, Richard Peng, Gabriela Pinto, Jinhu Qi, Ameen Qureshi, Namratha Sairam, Tanishq Salkar, Srivarshan Selvaraj, Kashish Atit Shah, Gokulraj Varatharajan, Reuben Varghese, Siyi Zhou, and Vito Zou.
\textbf{Previous memos:} \cite{memo1,memo3,memo4,memo5,memo6,memo7,memo8,memo9}

% \newpage

\bibliographystyle{ACM-Reference-Format}
\bibliography{chatgpt,genai}

\end{document}